# Designing technology, developing theory.
# Towards a symmetrical approach

Cornelius Schubert (cornelius.schubert@uni-siegen.de)
Andreas Kolb (andreas.kolb@uni-siegen.de)
Manuscript version 10th June 2020
Accepted for publication by Science, Technology, & Human Values

Abstract

We focus on collaborative activities that engage computer graphics designers and social scientists in systems design processes. Our conceptual symmetrical account of technology design and theory development is elaborated as a mode of mutual engagement occurring in an interdisciplinary trading zone, where neither discipline is placed at the service of the other, and nor do disciplinary boundaries dissolve. To this end, we draw on analyses of mutual engagements between computer and social scientists stemming from the fields of computer-supported cooperative work (CSCW), human–computer interaction (HCI), and science and technology studies (STS). We especially build on theoretical work in STS concerning information technology (IT) in health care and extend recent contributions from STS with respect to the modes of engagement and trading zones between computer and social sciences. We conceive participative digital systems design as a form of inquiry for the analysis of cooperative work settings, particularly when social science becomes part of design processes. We illustrate our conceptual approach using data from an interdisciplinary project involving computer graphics designers, sociologists, and neurosurgeons with the aim of developing patient-centered visualizations for clinical cooperation on a hospital ward.

Introduction

Software design and sociological analysis have come together at an interdisciplinary cross-roads since the 1980s. In this paper, we trace some of the enduring motifs and divisions of labor that have emerged from interactions between these two fields and discuss the respective "modes of engagement" (Ribes and Baker 2007) that link social scientists, computer scientists, and members of user communities. We sketch our own perspective of a symmetrical mode of engagement as a contribution to the current discussion in science and technology studies (STS) on "trading zones" between STS and technological design (Vertesi et al. 2017).

Our perspective draws on previous research that successfully engages both social scientists and computer scientists in human–computer interaction (HCI) and computer-supported cooperative work (CSCW). These approaches place a premium on ethnographic methods for the study of technology in use, on the conception of work as a fundamentally practical and mediated activity, and on user involvement in the design process (Bannon and Schmidt 1991; Robinson 1990). Moreover, they strive for a close collaboration between the fields of social science and computer science (Dourish and Button 1998), but are marked by a constant tension between ethnographic methods and software design (cf. Suchman 1994; Blomberg and Karasti 2013).

In line with the notion of a "trading zone" (Galison 1997), our symmetrical approach does not aim for a fusion of the disciplines but rather upholds disciplinary distinctions and configures the relations between social and computer science in two dimensions: First, on a conceptual level, it builds on a symmetrical understanding of the social and material aspects of designing technologies in STS (Callon 1987). It does not pit a technology-centric perspective against a user-centric approach, but focuses on the transformative capacity of technologies in real-life work settings such as IT in health-care (Berg 1998). Second, on an interdisciplinary level, it not only introduces STS concepts and methods into the design of IT systems, but also seeks to allow for the reverse effect, that is, the capacity of technological design to influence STS concepts (Latour 1996). We view this allowance as a necessary symmetry, as there are growing overlaps between designing technology and developing theory in research styles that advocate a continuous iterative engagement with empirical and conceptual results (Bryant 2017, 299–315; Rohde et al. 2017). In addition to these direct contributions to and benefits from technical design, STS also provides an analytic framework for setting up interdisciplinary collaborations in terms of the symmetrical approach, for instance by understanding the design of the prototype as heterogeneous problem solving using boundary objects (Star 1989).

Our empirical case is based on an interdisciplinary project involving computer graphics, neurosurgery, and sociology, which was established with the aim of developing patient-centered visualizations using an anatomical avatar for representing medical data. Within the current daily routine of the hospital, these data may not be circulated due to shift rotations and understaffing or they might be too time-consuming to retrieve from the distributed paperwork and hospital information systems. The project is set up in a way that each discipline contributes in the interdisciplinary research while at the same time seeking to extract an original disciplinary added value (cf. Spiller et al. 2015).

The authors of this paper are the principal investigators of the project. Cornelius Schubert is a sociologist specializing in STS and medical sociology; Andreas Kolb is a computer scientist specializing in computer graphics. We will elaborate how our symmetrical approach can be conceived as a mode of engagement and a trading zone between the different parties, and how their collaborative activities might be organized on the level of concrete research practices.

We present our argument as follows: The first part of the paper revisits the collaborative relations between technological design and sociological research that have emerged since the 1980s, taking the fields of HCI, CSCW, and STS as prominent examples. We then specify our symmetrical approach as a distinct mode of engagement and a trading zone in which social science, computer science, and user communities come together. In the third section, we briefly illustrate this approach using empirical data from our interdisciplinary project.

## Technological design and social theory in social and computer sciences: HCI, CSCW, and STS

Confronted with widespread problems in company IT systems in the 1980s, software engineers turned to the social sciences for help in designing systems that would better reflect the intricacies of collaborative work (Sommerville et al. 1993). Nascent fields such as HCI (Robinson 1990) and CSCW (Bannon and Schmidt 1991) drew heavily on findings from interactionist sociology, ethnomethodology, and anthropology. And the then-nascent field of STS showed a related interest in studying technological designs in order to fortify sociological theory (Callon 1987).

### HCI & CSCW

Robinson (1990) depicts the situation from a software engineering perspective in the field of HCI. Computer interfaces, he asserts, often demarcate the technical and the social aspects of software engineering, i.e., between "technical issues" and "non-technical issues" (ibid., 45).

In addition to this distinction, the then-prevalent understanding of human-computer interaction in software engineering was defined by simple cognitive models. According to Robinson, these dominant concepts ultimately led to a dual deficiency in software engineering, because the importance and the particulars of in situ human–machine interaction were widely disregarded. Robinson therefore argued for a stronger engagement with ethnographic methods on the part of software engineers.

Like HCI, CSCW calls upon sociology to enrich the engineer's understanding of work practices. Bannon and Schmidt (1991), for example, point out that in order to develop useful groupware applications, software designers need to understand the basic cooperative, i.e. social, features of work. They draw on concepts from industrial sociology as well as an interactionist sociology of work in order to highlight two main features of the sociality of work: First, cooperation mostly happens in a shared information space. Teams in industrial settings have a shared physical location and shared knowledge of the tasks at hand. Second, their cooperation cannot be completely planned in advance. There is always a residual need for situated adaptation and ad-hoc activities and coordination.

In the 1990s, an increasing mutual interest in understanding and designing information systems in organizational and work settings can be observed among both systems designers and social scientists (cf. Greenbaum and Kyng 1991; Orlikowski et al. 1996). However, crossing the "great divide" (Bowker et al. 1997) between social and computer science turned out to be more difficult than simply applying ethnographic methods and social theory to engineering problems or using technology design as an easily accessible field site. Given these firmly entrenched camps, critical accounts began to arise, questioning the dominant modes of engagement between social and computer science.

One main issue of contestation concerns the use and potential of ethnography as an empirical research method (Suchman 1994). Anderson (1994) asserted that systems design largely misconceived ethnography as an "impressionistic genre of reportage" (ibid. 178), conducted in order to gather "touchy-feely" data from the field, leaving the true analytic potential of ethnography largely untapped. This critique continues today, when reducing ethnography to a handful of "implications for design" (Dourish 2006) severely curtails its rich and detailed knowledge of the field and facilitates only "weak connections" (Blomberg and Karasti 2013, 395) with design. Emancipating ethnographic methods from design requirements is important in order to avoid the risk of neglecting elaborations by ethnographers that contain relevant conceptual insights into cooperative work practices and design issues (Schmidt 2011).

In sum, the modes of engagement between social and computer science in HCI and CSCW show a distinct pattern. First, they are characterized by strong theoretical and methodological input from social science into the field of systems design. Second, the growth in mutual interests between the disciplines can lead to fruitful collaborations and the formation of interdisciplinary approaches to design. However, as social science concepts and methods become increasingly incorporated into technological designs, they risk becoming mere "features" in service of computer science and losing their original descriptive and analytic potential. This does not lessen the fruitful and productive work of HCI and CSCW, but it does point to a unilateral mode of engagement where ideas tend to flow more heavily from social science to computer science than the other way around.

Despite close interdisciplinary engagements in fields such as HCI and CSCW, other areas of computer science do not share such established trading zones with the social sciences. The field of visualization and computer graphics, on which we draw in our case, is predominantly governed by technology-driven innovations. When integrating user perspectives in novel designs, the primary methods are evaluation techniques that attempt to measure the efficiency and efficacy of new (visualization) technologies with respect to a) data analysis and reasoning, b) communication through visualization, c) collaborative data analysis, or d) user performance (Lam et al. 2012). In practice, many user-related research projects in visualization and computer graphics address well-known problems for which quantitative measures of success and benchmarks already exist. In such cases, new technologies are evaluated without much active user involvement. Research projects that address more specific application scenarios are typically confronted with the additional requirement to define (quantitative) measures of success and benchmarks. Subsequently, projects such as ours are more difficult to "sell" to the visualization and computer graphics community, as it is often quite challenging and sometimes impossible to reconcile the requirements of quantitative benchmarks with the in situ design evaluations produced by ethnographic fieldwork and qualitative interviews.

## STS

Interest in the design and impact of technology in STS in the 1980s mirrors the growing interest of computer science in social science concepts and methods at the time. Callon (1987) argues that engineering projects do not begin with technical questions and gradually acquire social, political, and economic dimensions as they progress. Rather, they are fundamentally heterogeneous and complex from the very start. This circumstance forbids the clean separation of technical and social issues in the study of technical designs. The symmetry[1] proposed by actor–network theory (ANT) as a solution to this separation allows Callon to trace hetero-geneous associations and to study the sociomateriality of "society in the making."

---

[1] To be precise, studying heterogeneous association in this context pertains more to the principle of "free association" than to "generalised symmetry" Callon (1986, 200).

Latour (1996) extends this line of reasoning to computerized work sites. According to Latour, the engineering and social sciences exerted a strong influence on concepts of rationality and modernity in the 1950s and 1960s, but came under fire in the 1970s for neglecting the human dimension (ibid., 300). Computerized work sites make it obvious that a one-sided approach, favoring either technological or social dynamics, is largely misplaced. Rather, they point to the need for "a complete redefinition of the divide between the two worlds" (ibid.).

Callon and Latour highlight this central symmetry; while technology design profits from social theory, it also benefits from the study of technological design, *presenting a distinct added value for both sides from mutual collaboration*. A second symmetrical aspect is that *design should not prioritize human or technical agency alone*. According to Berg (1998), one main corollary of critique directed at technocratic concepts of design in HCI and CSCW is a tendency to lean into human-centric approaches and, in so doing, to neglect the transformative role of technology in social practices (cf. Jensen 2001; Garrety and Badham 2004). Prioritizing the human side of design while associating technology with negative concepts of authority and control tends to result in technological designs that are configured for the minimally invasive support of existing work practices, yet it simultaneously prevents disruptive, albeit potentially positive, technological change (Berg 1998, 469).

This brief account of STS perspectives related to the design of information systems sets the stage for our own symmetrical approach. First, we assume that *sociological concepts may themselves be transformed through the close study of technological design*, especially if they become actively involved in the design process (Rogers 1997). Second, the practice of *maintaining a difference between the social and the technical should be exploited for its analytic* *potential and not used as a basis for normative claims* (Jensen 2001). Both social and computer scientists then acknowledge that the social and the technological cannot be fundamentally separated in design processes, just as the design process should not pit a technology-centric perspective against a user-centric approach. In the tradition of STS, it seeks to navigate a course that stays clear of either social or technological determinisms while at the same time allowing technical design to be more mindful of social practices and social research to acknowledge the intricacies of technical design.

Collaborations of STS scholars with practitioners in fields that do not have a long-standing history of mutual involvement, such as computer graphics, may benefit from such a symmetrical perspective. As a mode of engagement, it offers a basis for the productive frictions of "agonistic-antagonistic" (Barry et al. 2008, 29) interdisciplinary endeavors that entail a mutual "letting go" (Spiller et al. 2015, 559–561) of disciplinary habits or comforts, while at the same time seeking out a disciplinary added value by crossing the "great divide" (Bowker et al. 1997) between computer and social science.

## Symmetrical positionings

The main feature of our symmetrical approach lies in the preservation of distinct disciplinary orientations that researchers and practitioners bring with them as they participate in close interdisciplinary collaborations. This approach heeds Bannon's call for a "strong" interdisciplinarity (1997, 372), where disciplines do not labor side by side but engage in fundamental exchanges based on their respective concepts and methods (cf. Dourish and Button 1998). Blomberg and Karasti have supported this view more recently with respect to ethnography and design (2013, 400-405) and we see our symmetrical approach in line with their position. In addition to the dominant mode of collaboration, or "bringing social theory into technological design" (Berg 1998), we specifically ask the inverse question, which is *how might we generate sociological insights from technological design processes*? The challenge is to create a mode of mutual engagement that does not place one discipline at the service of the other, nor create a synthesis that does away with disciplinary boundaries (Barry et al. 2008). This mode of engagement must align two seemingly contradictory objectives: (a) a distinct disciplinary added value derived from (b) close interdisciplinary collaboration on issues that cannot be addressed in a mono-disciplinary fashion.

**Modes of engagement and trading zones**

Ribes and Baker reflect on their involvement in the design of large information infrastructures to point out four elements that mark different modes of engagement: *development timeline*, *initiation*, *participation type*, and *details of involvement* (Ribes and Baker 2007, 111–114). Similar to Strathern (2004), they want to move beyond a "response mode" of social-research-as-a-service that is often called upon to address social issues surrounding technical designs. In the experience of Ribes and Baker, depending on the development timeline of a project, the time of initiation, and how social researchers are invited to participate, as well as the details of their involvement, social science can contribute much more than social fixes to technical problems. The authors note that deep and early involvement is often favorable when it comes to integrating original social science input (ibid. 113). In order to ensure robust interdisciplinary exchanges in our own symmetrical approach, we propose early mutual engagement as a vital element, together with strong participation and in-depth involvement of social scientists in the design process.[2] A symmetrical approach entails collaboration, in which all researchers are on equal footing by constituting a "trading zone" (Galison 1997). Building on the modes of engagement described by Ribes and Baker, Vertesi et al. (2017) specifically discuss the role of STS in the design of new technologies. They identify their own four modes of engagement: *corporate*, *critical*, *inventive*, or *using design as inquiry*. Each mode of engagement articulates a specific type of interdisciplinary collaboration occurring in distinct "trading zones" (ibid. 169). Of the four modes of engagement, "using design as inquiry" describes involvement on the part of social researchers in the design process that goes beyond analysis, critique, or constructive input to engage directly in configuring designs (ibid. 179–181). Symmetrical involvement of social sciences then extends past design implications or actively involving users. Instead, it pushes back on design issues, on sociological theory, and on research methods. Here, the notion of a trading zone is very useful, because specifying diverse modes of engagement is not enough to describe the manifold situated activities happening in such trading zones. Therefore, by building upon the concept of trading zones, we conceive of software design not as a linear progression from idea to artifact but as an iterative and recursive process that cycles back and forth between users, empirical research, theoretical and technological abstractions, design mock-ups and prototypes, until a more or less durable sociomaterial form is crafted (Suchman et al. 2002).

---

[2] A notable difference is that Ribes and Baker reflect on large-scale infrastructures whereas our focus is on cooperation in small teams.

## Contours of symmetry

Our discussion so far highlights the multiple modes of engagement and diverse trading zones that have evolved between social science and computer science. We note that there is little consensus on how technological design and sociological research should be related. We also emphasize that our approach is not meant to be universal. Rather, we view it as one option among others for articulating heterogeneous interests in interdisciplinary settings (Star 1993). Our project started in 2016 and is part of a larger collaborative research program on cooperative media at the University of Siegen with up to 12 years of funding. We focus on a specific mode of engagement and the constitution of an associated trading zone that encourage a symmetrical collaboration on several levels and prevent the different parties involved from working side-by-side or as a service to one another. The following six aspects form the basis of our symmetrical approach.

First, we *avoid the biases of either technology-centered or human-centered design* (Berg 1998) by conceiving of work practices as inherently sociomaterial. We also adhere to the principle of "free association" (Callon 1986, 200) by privileging neither humans nor technology in the accomplishment of complex distributed tasks. Following this basic principle, we combine the transformative benefits of technological design with a keen sensibility towards practical working relations. Symmetrical approaches drawing on ANT lend themselves to the analysis of information systems (Hanseth et al. 2004; Randall et al. 2007, 104–109) and recent approaches like "sociomaterial design" (Bjørn and Østerlund 2014) echo this need for an impartial perspective.

Second, we *acknowledge disciplinary discontinuities* between social and computer science and seek productive ways to exploit the "great divide" (Bowker et al. 1997). We aim to establish a trading zone between social and computer science that enables the development of individual disciplinary insights. Such disciplinary contributions, however, require close interdisciplinary collaboration on issues that cannot be addressed in mono-disciplinary fashion. Following Barry et al. (2008), we aim to create agonistic-antagonistic engagements that hold both disciplines in a productive, symmetrical tension.

Third, we enable researchers to *push back on their own disciplines*, for example, by critically questioning sociological theory based on the study of computerized work sites (Latour 1996; Dourish 2006) or by questioning technical design procedures based on observations gained through ethnographic research (Schmidt 2011, 355; Vertesi et al. 2017, 179–181). Close participation of users should also be provided for, in order to counterbalance the potential dominance of academic research (e.g. social or computer science). However, users may themselves be unaware of features of their cooperative work relations or of possible technological solutions (Sommerville et al. 1993).

Fourth, we conceive of research in the social and computer sciences, as well as collaborative work practices as *variations of a general type of problem solving*, specifically as processes of "inquiry" (Dewey 1938). The pragmatist notion of inquiry rests upon the idea that practical, common sense inquiry into daily life as well as more systematic processes of scientific inquiry are not inherently different.

Fifth, we propose that the design of technical systems and the development of sociological theory can be treated in a symmetrical manner under the condition that *the research styles employed by both social and computer scientists advocate a continuous, iterative engagement with empirical and conceptual results*. The compatibility of grounded theory with methodologies applied in software design has already been emphasized elsewhere (Bryant 2017, 299-315; Rohde et al. 2017). Seen through this lens, designing technology and developing theory both connect empirical investigations with evolving conceptual abstractions in order to categorize, classify, or model social realities.

Sixth, and lastly, we encourage *high-frequency and long-term collaboration* by all parties. The majority of tasks are not temporally or functionally divided into separate units but run parallel throughout all phases of research. This translates into a collaborative research practice in which the computer scientists take part in participant observations, organizing and conducting interviews, and analyzing field notes and transcripts. Likewise, the social scientists participate in selecting design features of the prototype, influencing the basic functional requirements for the user interface and the underlying software architecture. However, close collaboration does not exclude specific disciplinary interests. Rather, it provides a central resource for reflecting one's own position. Users are also constantly engaged in the design process through the discussion of mock-ups and prototypes, thereby balancing inputs from social and computer scientists with their practical work requirements.

Despite diverse attempts to bring software engineering and sociological research closer together, such approaches are still rare in both disciplines (Baxter and Sommerville 2011; Jackson et al. 2014). In the next section, we sketch out how such an approach can be put to work.

## The symmetrical approach in practice

We draw on a long-term collaborative research project that started in 2016 and is expected to run for 8 to 12 years within the Collaborative Research Centre "Media of Cooperation" funded by the German Research Foundation. The team[3] consists of two principal investigators, two PhD researchers (computer graphics and sociology), a part-time post-doc from media studies, and student assistants. The project cooperates with the neurosurgical ward[4] of a local, medium-sized, non-university, non-teaching hospital, where one neurosurgeon is partly funded by the project to facilitate ongoing consultation.

---

[3] Andreas Kolb, Julia Kurz, Dmitri Presnov, Cornelius Schubert, and Judith Willkomm.

Cornelius Schubert is a sociologist specializing in STS and medical sociology and has been conducting research at the intersections of medicine and IT since the early 2000s. He has collaborated with computer scientists on cooperative clinical work in two previous projects. Andreas Kolb is a computer scientist specializing in computer graphics and visualization. He has conducted one project in medical visualization in collaboration with a neurosurgical ward. Both principal investigators developed the grant proposal over roughly two years, focusing on a hybrid design approach that would integrate technology-driven, top-down methods with user-centered bottom-up approaches. Computer graphics is oriented towards technical advances and the aim within the project is to develop novel modes of visualizing clinical data. For the sociologists, the prime research motive is to gain deeper insight into the relations of visualization and cooperation and the material-semiotic constitution of cooperative clinical work. Over the course of the project, the idea of a symmetrical approach emerged, as it became clear that the respective research styles were sufficiently similar to attempt synchronization in ongoing collaboration. This especially pertains to the two PhD researchers, who collaborated intensively in gathering and analyzing empirical data and designing a prototype.

One crucial aspect of our symmetrical approach lies in configuring a mode of engagement that facilitates a disciplinary added value through high-frequency and long-term interdisciplinary collaboration. It conceives of technological design as a resource for sociological inquiry and therefore as a specific way of knowing (cf. Vertesi et al. 2017, 179) that offers unique insights into the material-semiotic relations of cooperative work settings. In our case, the added value pertains largely to the disciplines of sociology and computer graphics, since we do not produce scientific medical insights for the field of neurosurgery.

---

[4] We especially thank Daniel Alt, Veit Braun, Johannes Dillmann, Robert Zilke, as well as the other staff and patients of the Diakonie Klinikum Jung-Stilling in Siegen.

## Research methodology

In essence, our work is simultaneously about transgressing and maintaining disciplinary boundaries in order to create a productive tension that may lead to new disciplinary insights. We thus synchronize our research and design practices as sequences and loops to allow for the correction or modification of prior stages due to new findings or problems arising in later stages of our research. Each stage is assigned to a primary discipline, e.g., field research is the domain of sociology while computer graphics is responsible for technology design. At the same time, we encourage, and even force, an increased amount of interaction between the disciplines.

On the one hand, this setup of distinct yet shared responsibilities enables a tight interweaving of empirical data on cooperative practices into the development of new technical components and interfaces. On the other hand, such an arrangement requires a sufficient level of mutual understanding among participating disciplines as well as a mutual reference point, e.g. the technological design, as a basis for collaboration (Henderson 1991).

The mutual engagement in the design process thus affected the empirical work in the project, which started with participant observations in 2016 and 2017. Both PhD researchers, the media studies post-doc, and a student assistant from computer science, spent a total of 22 days observing cooperation among the neurosurgeons. The sociological PhD researcher spent 6 additional days observing cooperation in nursing care on the same ward. Building on these observations, paper mock-up interviews[5] with all the neurosurgeons were conducted in 2017 (8 group interviews) and elaborated into digital mock-up workshops with selected neurosurgeons in 2018 and 2019 (11 sessions). Finally, group evaluations of the prototype were conducted with the neurosurgical staff at the end of 2019 (7 group sessions). The mock-up sessions were collaboratively organized by both PhD researchers and recorded on video, where sociologists designed the interview and workshop setups, while computer scientists focused on the mock-ups featuring fundamental visualization concepts. Accompanying interviews were conducted with the nursing staff but did not enter the prototype design so far. Extending the prototype to nursing cooperation is planned for the next phase of the project.

Putting our symmetrical approach into practice rests on three components: (1) achieving a *shared understanding* requires establishing a trading zone for collaboration and, at least to some extent, a shared will to understand the respective goals and requirements of fellow collaborators; (2) *mutual involvement* is necessary so that each researcher may engage with the methods and concepts of other collaborators, creating a strong impulse for reflection of one's own work and that of others and (3) *common tools* facilitate the translation of empirical findings into novel technological and theoretical concepts by representing data in a mutually accessible format.

Achieving a *shared understanding* required open-minded and sufficient time. This especially applied to the time-consuming discussions in which implicit assumptions, ambiguities, and inaccuracies came to light and trust needed to be built (Dunker 2001). In our project, this was primarily done through regular team meetings in which the overall progress and direction were discussed and additional meetings focused on specific issues. The PhD researchers met nearly every week, whereas the principal investigators met with the team roughly once a month, sometimes also bi-weekly. Fostering *mutual involvement* is clearly more challenging, as it required actively integrating perspectives from different disciplines into one's own research activities (Spiller et al. 2015). In our project, this was accomplished through collaborative fieldwork and design, which at the same time are sources of constant tension, since all participants have to leave the comfort zone of their own discipline. On the level of *common tools*, we drew on visual notations commonly used in software engineering to bring together our varying perspectives (cf. Ensmenger 2016 for the collaborative use of flowcharts). Unified Modeling Language (UML) (see Booch et al. 2005) was one of the tools employed to link sociological analysis with technical design.

---

[5] The paper mock-ups consisted of early visualization sketches printed on cardboard paper that were handed out and discussed in the interviews.

In the following, we concentrate on UML as a common tool. In line with Henderson (1991, 450), we understand UML in our project as a boundary object that is wielded in order to "socially organize distributed cognition" and to allow "members of different groups to read different meanings particular to their needs from the same material." Among the many exchanges occurring in our trading zone, UML allows for a level of abstraction that sits between detailed ethnographic descriptions of cooperative practice and its algorithmic representations in software (cf. van Dyke Parunak and Odell 2002; Trkman and Trkman 2014). The use of UML was advocated by computer graphics in order to discuss the basic requirements of the system with respect to the cooperative tasks on the ward. However, this did not imply UML as a ready-made interface between social and computer science. Rather, it was employed as a means of organizing mutual exchanges by structuring the basic components of the system.

**UML – modeling as collaborative practice**

UML is a "general-purpose, developmental, modeling language in the field of software engineering that is intended to provide a standard way to visualize the design of a system" (Booch et al. 2005). Such tools are essential to any software development process where the main challenge lies in the translation of a non-trivial, real-world problem into a computer program. UML helps to create software models for the implementation of the concrete system. As such, the software models need to represent all information considered relevant about the real-world problem's structure, the involved object types, and their behavior in a consistent set of rules. By implication, the model describes the state and representation space of the program and thus its ability to correctly reflect the real-world situation and the solution of the related problem. Therefore, the disambiguation of linguistic inaccuracies and practical indeterminacies, which are every day and often implicit aspects of communication and work that need to be explicitly resolved in systems design, is a major task in software modeling.

Typically, the visual notations of UML are used to create diagrammatic representations of selected issues. For instance, we focused on four diagram types that allow for an acceptable trade-off between the complexity of real-world problems and the formal requirements of a software model: the use-case diagram, the class diagram, the component diagram, and the sequence diagram. Of these four, we will highlight the use-case diagram, because it is the starting point for the software modeling of real-world problems, followed by a further refinement of the design. Use-case diagrams represent a user's interaction with the software system and demonstrate the relationship between one or several users and different use case scenarios. We employ this diagram type to identify the parties involved (users, e.g. physicians and nurses) and the usage scenarios (use-cases, e.g. checking the patient's status or the course of treatment).

More specifically, the use-case diagram serves as an interface where insights from ethnographic observations, interviews, and group meetings can be transferred (in part) into the architecture of the software system under development. It constitutes a representational space in which sociology and computer graphics negotiate the relations of cooperative medical work and collaborative software design, i.e. where the "double tuning process" (Jensen 2001, 213) between social practices and technical systems takes place. As formal notation, UML forces the reduction of empirical and conceptual ambiguities to computable classifications. For instance, terms like "interaction," "code," and "model" carry different meanings in social and computer science and need to be clarified. This often requires stricter categorizations than those created through qualitative analysis, yet both social and computer science share an interest in creating meaningful abstractions from empirical data and both can be seen as ways of modeling social realities (Bryant 2017, 299–315). At first sight, this challenges sociological reasoning to conform with computational requirements; at the same time, it forces software design to acknowledge the messy social realities that lie outside the confines of formal modes of data acquisition. In line with our symmetrical understanding, UML and the use-case diagram constitute transitional zones in which sociological research informs systems design (Berg 1998), just as systems design offers the potential to re-configure sociological concepts (Rogers 1997). This is by no means a smooth integration of sociological data into computer code but rather a long and thorny process of negotiation. Both sociology and computer science thus have to generate sufficient "interactional expertise" to facilitate the exchanges in a trading zone that supports this potential (cf. Gorman 2010).

**Constructing a use case of medical cooperation**

Our empirical research on the neurosurgical ward focuses on clinical cooperation taking place before and after surgery. This consists of the initial reception of the patient, the acquisition of relevant medical data, additional diagnostics, surgery preparation, the post-operative acquisition of medical data, and discharge. Besides neurosurgeons, this also involves nurses and sometimes technicians. However, due to shift rotations and understaffing, relevant patient information might not be passed on to the next shift verbally or might be too time-consuming to retrieve from the circulating paperwork and stationary information systems. As one neurosurgeon noted for shift handovers in an interview: "They are just about what I have to do right away, so it gets done. This is the only useful information that gets passed on. The rest, the condition of the patient, what was done, I can't remember all of this in half an hour. That's how it is. This is why I rather need a good documentation that I can quickly look up myself if something needs to be done and then I have all the information." (I1, 58:27, transl. CS). We address this problem by designing new visual modalities for representing cooperatively used clinical data.

Yet the sociology of organized medicine has, time and again, shown how work practices resist attempts at formalization and standardization (Timmermans and Berg 1997; Bruni 2005). Every design of a technical system in this context is bound to face thorny dilemmas between articulations of practical work, the elicitation of system requirements, and representations of cooperation in abstract models (Gerson and Star 1986, 258). A second neurosurgeon underscored this dilemma from his perspective: "I have difficulties to … humans are very complex. And in addition, we have the individuality of every patient and I have difficulties fitting them into an objective grid. And at the same time, I require this grid to be complete for improved information transmission." (I3a, 47:24, transl. CS).

Our use of UML reflects these inherent tensions. Its formal notations require all parties to negotiate how relevant features of clinical cooperation might be translated into visual representations. This does not entail a full formalization of all cooperative aspects. Rather, participating actors are required to maneuver delicately between the practical ambiguities of medical knowledge, formal and informal organizational procedures, possible visual modalities, and user feedback. In line with Berg and Toussaint (2003, 228), we emphasize that (UML) models will always be partial and that this partiality is a central issue when re-integrating the resulting visualization system into the daily procedures of the hospital ward by allowing flexible access to relevant information, along with options to support aggregation and linkages. We also bear in mind that designing technology in our case cannot be separated from interfering with social practices. Indeed, this is one of the main demands on the part of neurosurgery teams, who want to reflect on and adjust their ways of working together. A third neurosurgeon commented on the problems of daily data retrieval in a paper mock-up interview: "Of course you can find the information. If you are careful, you can find it written down somewhere in the computer, or the record, or the chart. But you have to read all of this to find the relevant information. And here [on the mock-up] you can see it at a glance, because you can select 'muscle strength' and then you see muscle and only muscle." (I7, 25:59, transl. CS). The following example illustrates how we relate sociological conceptualization and technical design through UML diagrams.

Figure 1 shows a section of a UML use-case diagram concerning the feature "display patient status" that a physician (A) may access in practice. After login and the selection of the patient, the system will display the patient status (B) in an aggregated fashion, i.e. depicting lesions or symptoms in an overview, e.g. as requested by the neurosurgeon in I7. From there, the user can change to an extended display format, such as accessing detailed patient data (C) or visualizing disease dynamics (D). The same neurosurgeon pointed out the need for this later in the interview: "Most times, you need the most relevant information now, while you are at the bedside examining the patient. You want to know now, if the patient takes aspirin or not. You want to know now, if the patient had this paralysis earlier or not." (I7, 32:33, transl. CS). Furthermore, accessing detailed patient data comprises two distinct data classes: Data having an anatomical reference that can be displayed on an avatar directly (E) and data without anatomical reference (F). This diagram was developed based on ethnographic observations and interviews with selected personnel; the results were then used to sketch out initial ideas about how the system can be used in clinical practice, i.e., to quickly access diagnostic data when seeing a patient for the first time or after several shift rotations.

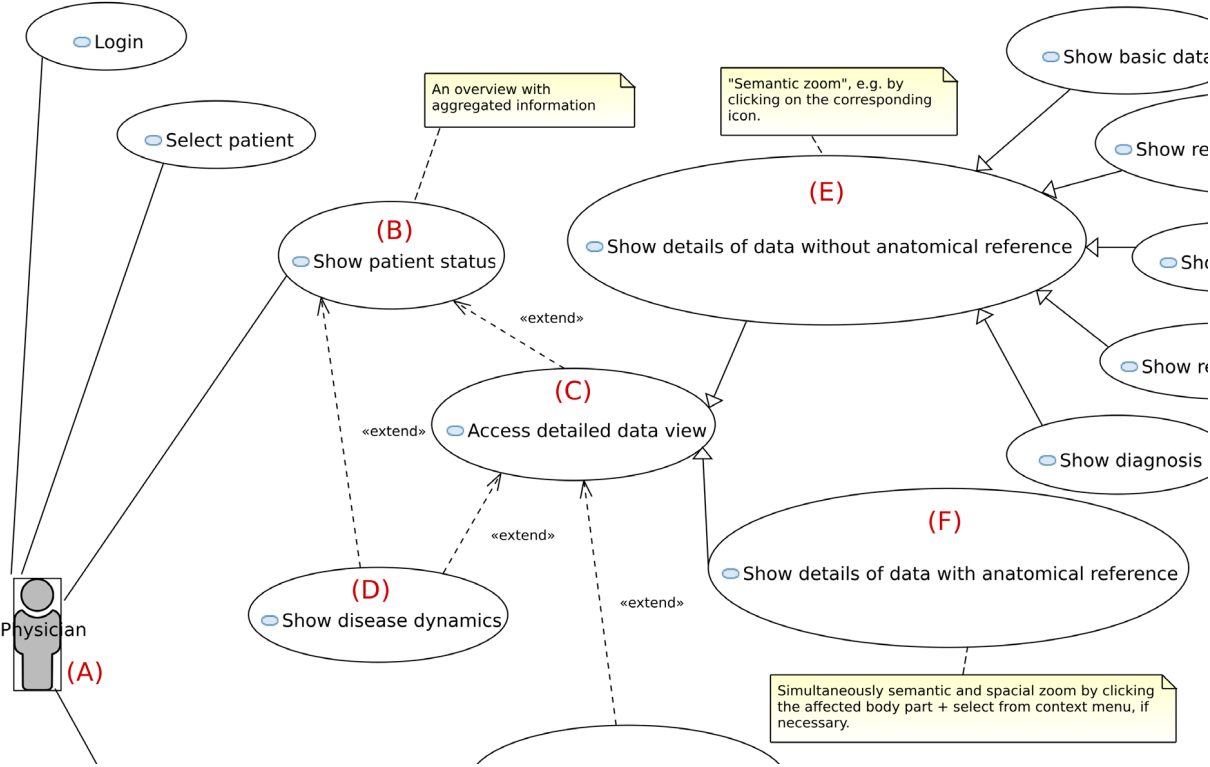

Figure 1. UML use case diagram section for feature: display patient status.

Constructing several UML use-cases eventually led the sociologists to reconsider their view of cooperation on the ward. Once several empirical scenarios (e.g., patient intake and follow-up examination) had been formalized into UML diagrams, their differences became less obvious. The emerging similarities across the uses-cases subsequently formed into identifying a dominant "diagnostic mode" that lies at the heart of most cooperative medical activities on the ward. When working in this mode, the neurosurgeons follow a basic format of diagnosing disorders. From the patient intake to surgery preparation and post-operative monitoring, their cooperative requirements result from the constant necessity of being aware of the patient's current status, especially in the absence of colleagues from previous shifts. The identification of the dominant diagnostic mode then became a crucial element for selecting and displaying relevant clinical data. This, in turn, called for significantly more flexibility from computer graphics in the development of the visualization prototype to efficiently incorporate new data types and visualization requirements. Our point is that this "double tuning process" (Jensen 2001, 213) of social research and technical design crystallized in the collaborative work on UML diagrams in the manner described above, because of the enabling and constraining features of UML as a common tool.

Constructing the use cases and reflecting on the possibilities for visualizing the patient's status on an anatomical avatar, we further concentrated on the somatic "locatability" of information as a distinctive characteristic. First, somatic locatability poses a relevant research problem for computer graphics, as it requires visualizing information in a defined and often limited space. Second, our fieldwork showed how heavily neurosurgeons rely on precise somatic locatability in their daily work. While information related to neural disorders is predominantly related to different skin regions, muscles, and tendons, associated information such as blood values or medication has no, or only very weak, ties to specific anatomic structures. Thus, neural disorders lend themselves well to patient-centered visualizations because they can be precisely mapped onto an anatomical avatar. Our decision to concentrate on data with concrete anatomical reference then increased the degree to which somatic locatability became a central criterion in the design process.

Representing the dominant diagnostic mode of somatic locatability also required refining how it would be visually displayed to users. Again, we used UML to create objects and classes that included anatomical structures, especially symptoms of spinal cord disorders. These disorders play an essential role in the diagnosis of neurosurgical conditions and include various types and intensities of pain and numbness related to specific skin regions (dermatomes) or the level of strength related to specific muscles (myotomes). A specific spinal disc herniation, for instance, has corresponding symptoms of pain or numbness on the related dermatome. Figure 2 shows how clinical data can be represented by somatic locatability on an anatomical avatar using visual modalities such as different colors and textures superimposed onto sections of the body to indicate different intensities of pain and numbness.

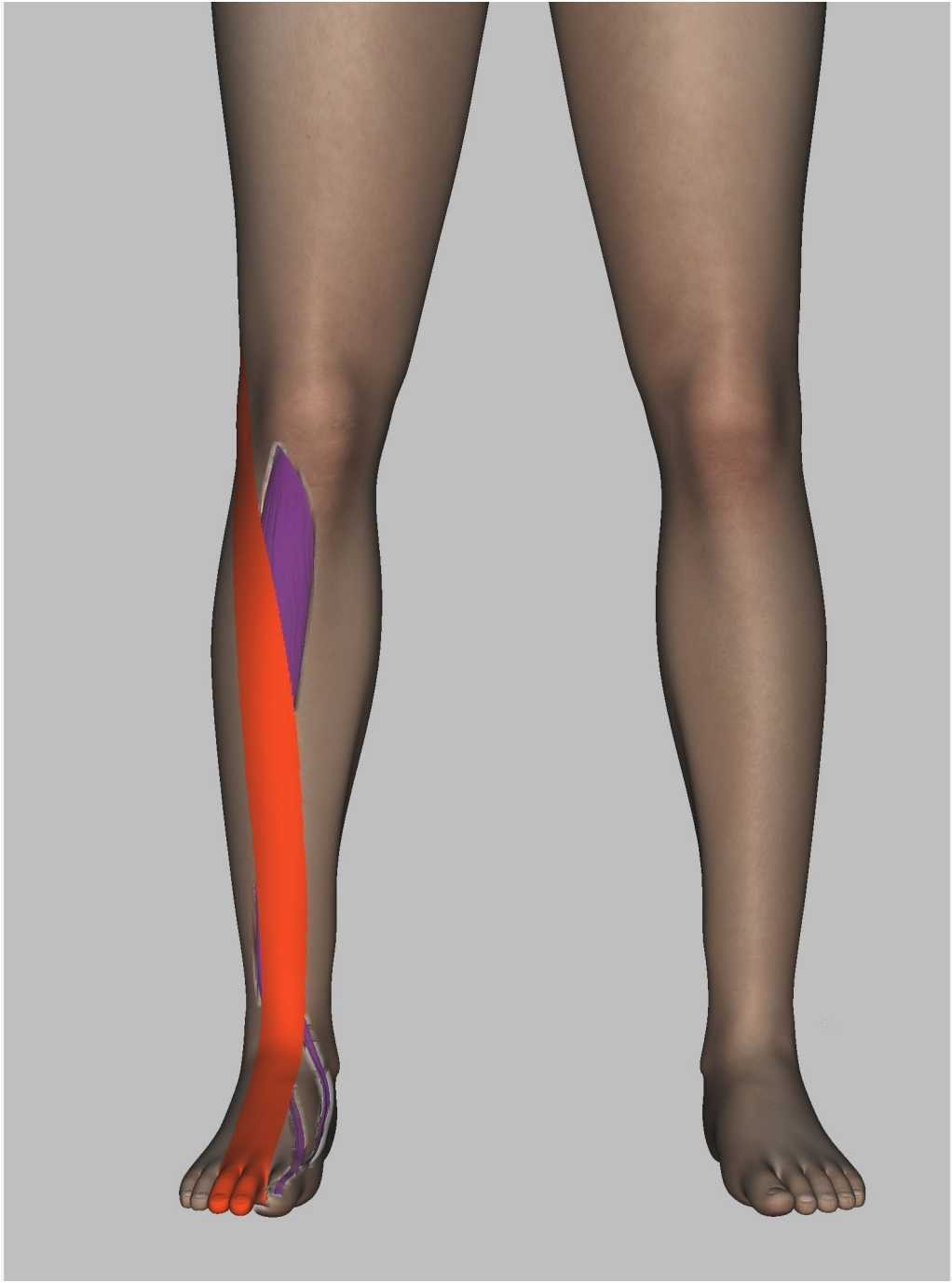

Figure 2. Superimposing symptoms of a spinal disc herniation on an anatomical avatar.

Sorting out the relevant disorders to be represented is one of the most crucial aspects of the design process, since fundamental decisions about the architecture of the system occur at this stage. As the visual modalities of color and texture only allow for the representation of two independent data items per location at a time, special care needs to be taken in information selection and grouping. This subsequently led us to reconsider the way information should be represented within the system, how data should be visualized using alternative modes such as time, and how such a system could be made productive within the heterogeneous information ecology of a hospital ward. While this analysis provided the core architecture for the computer graphics visualizations, the sociologists then returned to their empirical data in order to take a closer look at the different cooperative media and data practices on the ward, from verbal face-to-face and telephone communication to hard-copy medical records and the digital hospital information system, and finally the ubiquitous post-it notes and printouts containing handwritten notes. Again, we see that a double tuning process occurred between social research and technical design.

The above examples illustrate how the use of UML within our symmetrical approach does not simply funnel sociological knowledge into technical design but rather enables mutual engagements between both disciplines. Of course, UML is not a neutral means to this end, as it was designed as a standardized tool for object-oriented programming. Nonetheless, we see some of its components as a productive trade-off between the complexity of clinical practice and the formal requirements of software engineering and thus as an important tool within our interdisciplinary trading zone. Designing technology and building theory in our case do not inhabit incommensurable spheres, rather they are tightly related in collaborative research practice.

# Discussion and conclusion

Our main aim in this paper was to outline a symmetrical mode of engagement between the disciplines of sociology and computer graphics. Our collaboration was conceived in terms of a trading zone, in which we converged on selected issues, while diverging on others. We illustrated this by using UML as an empirical example of how partial convergence might be achieved, while at the same time allowing for distinct disciplinary impulses. At its core, the symmetrical approach seeks to avoid unilateral flows of social theory into technical design, (and, vice versa, of computational solutions into social practices) and instead to create an unfettered interdisciplinary exchange while avoiding the pitfalls of social and technical determinisms. We argued that this approach is particularly suited to collaborations in which both disciplines rely on research styles that continuously and iteratively engage with empirical and conceptual findings, as it is the case, at least for some fields, in social and computer science. This collaboration is neither seamless nor without strain; rather it is marked by high-frequency and long-term exchanges and an openness on both sides to (participating in) the research practices of the other.

Our symmetrical approach seeks to enable disruptive technology development on the part of computer graphics experts, i.e., to support the development of new modalities for visualizing clinical data, while at the same time remaining sensitive to concrete work practices and their resistance to change. It does so by carefully fitting the prototype to the existing information ecology of the ward, including vocal interaction, formal and informal paper-based records, as well as stationary and mobile digital devices. The sociological benefit lies in the close participation in software design processes, an involvement that essentially shapes a uniquely constructive digital sociology (Goulden et al. 2017). This process led to the identification of a dominant diagnostic mode in clinical cooperation and to the reconsideration of cooperative media on the ward. In sum, the symmetrical approach enables the sociologists to see the social through the technical (i.e. through the design lens of a technical system) while allowing computer graphics specialists to perceive the technical through the social (i.e. the negotiation of technical features vis-à-vis collaborative work practices).

We add to the growing discussion of digitalSTS (Vertesi and Ribes 2019), especially to the benefits and challenges of becoming involved in technical design. We argue that becoming part of a design process can be a unique occasion for reflection and inquiry as well as a way of pushing back on sociological analysis and concepts. The challenges for STS lie not only in locating the politics within technical design but also in becoming part of the process and having to situate novel digital devices within existing ecologies of IT, paper, and interaction. This close-up and hands-on involvement with digital design and digital practices provides a grounded approach to processes of digitization and it offers some new facets to building theory.

As the social sciences and STS are increasingly drawn into the design of digital technologies, their modes of engagement and the trading zones between them require closer inspection. By outlining our symmetrical approach, we hope to contribute to an increased mutual understanding between social and technical disciplines and benefit from both sides of the presumed, and perhaps slowly shrinking, great divide.

The research project is funded by the German Research Foundation through the Collaborative Research Centre 1187 "Media of Cooperation" at the University of Siegen.

# References

Anderson, Robert J. 1994. Representations and requirements: The value of ethnography in system design. *Human–Computer Interaction* 9 (2): 151–182.

Bannon, Liam J. 1997. Dwelling in the "Great Divide": The case of HCI and CSCW. In *Social science, technical systems, and cooperative work: Beyond the great divide*, eds. Geoffrey C. Bowker, Susan L. Star, Les Gasser and William Turner, 355–377. Mahwah: Lawrence Erlbaum.

Bannon, Liam J., and Kjeld Schmidt. 1991. CSCW: Four characters in search of a context. In *Studies in Computer Supported Cooperative Work. Theory, Practice and Design*, eds. John M. Bowers and Stephen D. Benford, 3–16. Amsterdam: North-Holland.

Barry, Andrew, Georgina Born, and Gisa Weszkalnys. 2008. Logics of interdisciplinarity. *Economy and Society* 37 (1): 20–49.

Baxter, Gordon, and Ian Sommerville. 2011. Socio-technical systems: From design methods to systems engineering. *Interacting with Computers* 23 (1): 4–17.

Berg, Marc. 1998. The politics of technology: On bringing social theory into technological design. *Science, Technology, & Human Values* 23 (4): 456–490.

Berg, Marc, and Pieter Toussaint. 2003. The mantra of modeling and the forgotten powers of paper: A sociotechnical view on the development of process-oriented ICT in health care. *International Journal of Medical Informatics* 69 (2): 223–234.

Bjørn, Pernille, and Carsten Østerlund. 2014. *Sociomaterial-Design: Bounding technologies in practice.* Cham: Springer.

Blomberg, Jeanette, and Helena Karasti. 2013. Reflections on 25 Years of ethnography in CSCW. *Computer Supported Cooperative Work (CSCW)* 22 (4-6): 373–423.

Booch, Grady, James Rumbaugh, and Ivar Jacobson. 2005. *The Unified Modeling Language user guide.* Reading: Addison-Wesley.

Bowker, Geoffrey C., Susan L. Star, Les Gasser, and William Turner (eds.). 1997. *Social science, technical systems, and cooperative work: Beyond the great divide.* Mahwah: Lawrence Erlbaum.

Bruni, Attila. 2005. Shadowing software and clinical records: On the ethnography of non-humans and heterogeneous contexts. *Organization* 12 (3): 357–378.

Bryant, Antony. 2017. *Grounded theory and grounded theorizing: Pragmatism in research practice.* Oxford: Oxford University Press.

Callon, Michel. 1986. Some elements of a sociology of translation: Domestication of the scallops and the fishermen of Saint Brieuc bay. In *Power, action and belief: a new sociology of knowledge?*, ed. John Law, 196–233. London: Routledge.

Callon, Michel. 1987. Society in the making: The study of technology as a tool for sociological analysis. In *The social construction of technological systems*, eds. Wiebe E. Bijker, Thomas P. Hughes and Trevor J. Pinch, 83–103. Cambridge: MIT Press.

Dewey, John. 1938. *Logic. The theory of inquiry*. New York: Henry Holt.

Dourish, Paul. 2006. Implications for design. In *Proceedings of the SIGCHI Conference on Human Factors in Computing Systems*, 541–550. SIGCHI conference, Montréal. New York: ACM Press.

Dourish, Paul, and Graham Button. 1998. On "technomethodology": Foundational relationship between ethnomethodology and system design. *Human–Computer Interaction* 13 (4): 395–432.

Duncker, Elke. 2001. Symbolic communication in multidisciplinary cooperations. *Science, Technology, & Human Values* 26 (3): 349–386.

Ensmenger, Nathan. 2016. The multiple meanings of a flowchart. *Information & Culture: A Journal of History* 51 (3): 321–351.

Galison, Peter. 1997. The trading zone: Coordinating action and belief. In *Image and logic: A material culture of microphysics*, 781–844. Chicago: University of Chicago Press.

Garrety, Karin, and Richard Badham. 2004. User-centered design and the normative politics of technology. *Science, Technology, & Human Values* 29 (2): 191–212.

Gerson, Elihu M., and Susan L. Star. 1986. Analyzing due process in the workplace. *ACM Transactions on Information Systems* 4 (3): 257–270.

Gorman, Michael E. (ed.). 2010. *Trading zones and interactional expertise: Creating new kinds of collaboration.* Cambridge: MIT Press.

Goulden, Murray, Christian Greiffenhagen, Jon Crowcroft, Derek McAuley, Richard Mortier, Milena Radenkovic, and Arjuna Sathiaseelan. 2017. Wild interdisciplinarity: Ethnography and computer science. *International Journal of Social Research Methodology* 20 (2): 137–150.

Greenbaum, Joan, and Morten Kyng (eds.). 1991. *Design at work: Cooperative design of computer systems.* Hillsdale: Lawrence Erlbaum.

Hanseth, Ole, Margunn Aanestad, and Marc Berg. 2004. Guest editors' introduction: Actor-network theory and information systems: What's so special? *Information Technology & People* 17 (2): 116–123.

Henderson, Kathryn. 1991. Flexible sketches and inflexible data bases: Visual communication, conscription devices, and boundary objects in design engineering. *Science, Technology, & Human Values* 16 (4): 448–473.

Jackson, Steven J., Tarleton Gillespie, and Sandy Payette. 2014. The policy knot: Re-integrating policy, practice and design in CSCW studies of social computing. In: *Proceedings of the 17th*

ACM Conference on Computer Supported Cooperative Work & Social Computing (CSCW 2014), 588–602, New York: ACM.

Jensen, Casper B. 2001. CSCW design reconceptualised through science studies. *AI & Society* 15 (3): 200–215.

Lam, Heidi, Enrico Bertini, Petra Isenberg, Catherine Plaisant, and Sheelagh Carpendale. 2012. Empirical studies in information visualization: Seven scenarios. *IEEE transactions on visualization and computer graphics* 18 (9): 1520–1536.

Latour, Bruno. 1996. Social theory and the study of computerized work sites. In *Information technology and changes in organizational work*, eds. Wanda J. Orlikowski, Geoff Walsham, Matthew R. Jones and Janice L. DeGross, 295–307. London: Chapman & Hall.

Orlikowski, Wanda J., Geoff Walsham, Matthew R. Jones, and Janice L. DeGross (eds.). 1996. *Information technology and changes in organizational work.* London: Chapman & Hall.

Randall, Dave, Richard Harper, and Mark Rouncefield. 2007. *Fieldwork for design.* London: Springer.

Ribes, David, and Karen Baker. 2007. Modes of social science engagement in community infrastructure design. In *Communities and technologies 2007: Proceedings of the third communities and technologies conference, Michigan State University 2007*, 107–130, London: Springer.

Robinson, Hugh. 1990. Towards a sociology of human-computer interaction. In *Computers and Conversation*, eds. Paul Luff, David M. Frohlich and Nigel Gilbert, 39–49. London: Academic Press.

Rogers, Yvonne. 1997. Reconfiguring the social scientist: Shifting from prescription to proactive research. In *Social science, technical systems, and cooperative work: Beyond the great* *divide*, eds. Geoffrey C. Bowker, Susan L. Star, Les Gasser and William Turner, 57–77. Mahwah: Lawrence Erlbaum.

Rohde, Markus, Peter Brödner, Gunnar Stevens, Matthias Betz, and Volker Wulf. 2017. Grounded Design – a praxeological IS research perspective. *Journal of Information Technology* 32 (2): 163–179.

Schmidt, Kjeld. 2011. *Cooperative work and coordinative practices: Contributions to the conceptual foundations of Computer-Supported Cooperative Work (CSCW).* London: Springer.

Sommerville, Ian, Tom Rodden, Peter Sawyer, and Richard Bentley. 1993. Sociologists can be surprisingly useful in interactive systems design. In *People and computers VII: Proceedings of HCI 92, York, September 1992*, eds. Andrew Monk, Dan Diaper and Michael D. Harrison, 342–354. Cambridge: Cambridge University Press.

Spiller, Keith, Kirstie Ball, Elizabeth Daniel, Sally Dibb, Maureen Meadows, and Ana Canhoto. 2015. Carnivalesque collaborations: Reflections on 'doing' multi-disciplinary research. *Qualitative Research* 15 (5): 551–567.

Star, Susan L. 1989. The structure of ill-structured solutions. Boundary objects and heterogeneous distributed problem solving. In *Distributed artificial intelligence. Volume II*, eds. Les Gasser and Michael N. Huhns, 37-54. London: Pitman.

Star, Susan L. 1993. Cooperation without consensus in scientific problem solving: Dynamics of closure in open systems. In *CSCW: Cooperation or Conflict?*, ed. Steve Easterbrook, 93–106. London: Springer.

Strathern, Marilyn. 2004. *Commons and borderlands: Working papers on interdisciplinarity, accountability and the flow of knowledge.* Wantage: Kingston.

Suchman, Lucy A. 1994. Working relations of technology production and use. *Computer Supported Cooperative Work (CSCW)* 2 (1-2): 21–39.

Suchman, Lucy A., Randall Trigg, and Jeanette Blomberg. 2002. Working artefacts: Ethnomethods of the prototype. *British Journal of Sociology* 53 (2): 163–179.

Timmermans, Stefan, and Marc Berg. 1997. Standardization in action: Achieving local universality through medical protocols. *Social Studies of Science* 27: 273–305.

Trkman, Marina, and Peter Trkman. 2014. Actors' misaligned interests to explain the low impact of an information system: A case study. *International Journal of Information Management* 34 (2): 296–307.

van Dyke Parunak, Henry, and James J. Odell. 2002. Representing Social Structures in UML. In *Agent-oriented software engineering II: Second international workshop*, ed. Michael J. Wooldridge, 1–16. Berlin: Springer.

Vertesi, Janet, and David Ribes (eds.). 2019. digitalSTS: A field guide for science & technology studies. Princeton: Princeton University Press.

Vertesi, Janet, David Ribes, Laura Forlano, Yanni Loukissas, and Marisa L. Cohn. 2017. Engaging, designing, and making digital systems. In *The handbook of science and technology studies*, ed. Ulrike Felt, Rayvon Fouché, Clark A. Miller and Laurel Smith-Doerr, 169–193. Cambridge: MIT Press.

Author Biographies

**Andreas Kolb** is the head of the Computer Graphics and Multimedia Systems Group, University of Siegen, Germany. He received his Diploma Ph.D. in Mathematics at the University of Erlangen, Germany, in 1992 and his Ph.D. in Computer Science with a focus in Computer Graphics at the same University in 1995. His research interests are computer graphics, visualization and computer vision.

**Cornelius Schubert** is a lecturer on innovation studies at the University of Siegen. He studied Sociology, Psychology and English at the University of Kassel and the University of Technology Berlin, where he received his Ph.D. in Sociology in 2006. He specializes in science and technology studies as well as medical and organizational sociology with a focus on ethnographic methods for studying human-technology relations.